\documentclass{aastex631}

\shorttitle{Detection of the SNR G312.4$-$0.4 with Fermi}
\shortauthors{P. Chambery et al.}
\graphicspath{{./}{figures/}}

\begin{document}

\title{\emph{Fermi}-LAT detection of the supernova remnant G312.4$-$0.4 in the vicinity of 4FGL J1409.1$-$6121e}

\correspondingauthor{Pauline Chambery, Marianne Lemoine-Goumard}
\email{chambery@lp2ib.in2p3.fr, lemoine@lp2ib.in2p3.fr}

\author{Pauline Chambery}
\affiliation{Universit\'e Bordeaux, CNRS, LP2I Bordeaux, UMR 5797, F-33170 Gradignan, France}

\author[0000-0002-4462-3686]{Marianne Lemoine-Goumard}
\affiliation{Universit\'e Bordeaux, CNRS, LP2I Bordeaux, UMR 5797, F-33170 Gradignan, France}

\author[0000-0002-6738-9351]{Armelle Jardin-Blicq}
\affiliation{Universit\'e Bordeaux, CNRS, LP2I Bordeaux, UMR 5797, F-33170 Gradignan, France}

\author[0000-0002-9238-7163]{Atreyee Sinha}
\affiliation{EMFTEL Department and IPARCOS, Universidad Complutense de Madrid, 28040 Madrid, Spain}

\author[0000-0001-9633-3165]{J.~Eagle}
\affiliation{NASA Goddard Space Flight Center, Greenbelt, MD 20771, USA}



\begin{abstract}

Gamma-ray emission provides constraints on the non-thermal radiation processes at play in astrophysical particle accelerators. This allows both the nature of accelerated particles and the maximum energy that they can reach to be determined. Notably, it remains an open question to what extent supernova remnants (SNRs) contribute to the sea of Galactic cosmic rays. In the Galactic plane, at around $312^{\circ}$ of Galactic longitude, \emph{Fermi}-LAT observations show an extended source (4FGL J1409.1$-$6121e) around five powerful pulsars. This source is described by one large disk of $0.7^{\circ}$ radius with a high significance of $45\sigma$ in the 4FGL-DR3 catalog. Using 14 years of  \emph{Fermi}-LAT observations, we revisited this region with a detailed spectro-morphological analysis in order to disentangle its underlying structure. Three sources have been distinguished, including the supernova remnant G312.4$-$0.4 whose gamma-ray emission correlates well with the shell observed at radio energies. The hard spectrum detected by the LAT, extending up to 100 GeV without any sign of cut-off, is well reproduced by a purely hadronic model.

\end{abstract}

\keywords{\emph{Fermi}, supernova remnant, G312.4$-$0.4}

\section{Introduction} \label{sec:intro}
Since the launch of the \emph{Fermi} Gamma-ray Space Telescope in June 2008, four catalogs\footnote{https://fermi.gsfc.nasa.gov/ssc/data/access/lat/} of sources detected with the Large Area Telescope (LAT) as well as 3 extensions of the fourth catalog \citep{Abdollahi_2020} have been released. All of them share a common characteristic: an important fraction of the sources ($\sim$1/3) are unidentified. This is especially true in the Galactic plane due to source confusion introduced by both the large density of sources and the prominent diffuse emission. The variability of the signal is certainly the most direct characteristic to associate a source with a binary system, a pulsar or an AGN. Pulsar wind nebulae (PWNe) and supernova remnants (SNRs), on the other hand, are not expected to be variable, so the association relies mainly on the spatial morphology: finding a coherent source extension across different energy bands can help to associate a \emph{Fermi}-LAT source with a potential counterpart. In addition, such multi-wavelength studies can also help to determine the emission mechanisms that produce the gamma-ray photons. This has already been done for several SNRs detected by the LAT \citep{2016ApJS..224....8A}.\\
Here, we focus on 4FGL J1409.1$-$6121e, one of the seven sources flagged as confused or contaminated by the diffuse background in the \emph{Fermi} Galactic Extended Source Catalog \citep{2017ApJ...843..139A}. This source was fitted with a uniform disk of $0.73^{\circ} \pm 0.02^{\circ} \pm 0.06^{\circ}$ while the cross-check with a Gaussian provided an extension of $0.51^{\circ} \pm 0.02^{\circ}$, showing that the fit of this extended source was not stable with respect to the assumed shape. This non-variable source\footnote{Its variability\_index of 4.2 in the 4FGL-DR3 Catalog is well below the threshold of 24.725 to claim a significant variability.} lies at (RA$_{\rm J2000}$, Dec$_{\rm J2000}$ = $212.29^{\circ}$, $-61.35^{\circ}$), in a 1.0$^{\circ}$ region containing five pulsars. These have high spin-down power and different ages which could power pulsar wind nebulae or gamma-ray halos (PSR~J1406$-$6121, PSR~J1410$-$6132, PSR~J1413$-$6205, PSR~J1412$-$6145, PSR~J1413$-$6141). These two last pulsars are coincident with the supernova remnant G312.4$-$0.4, discovered by the Molonglo Observatory Synthesis Telescope (MOST) at 408 MHz \citep{1985MNRAS.216..753C}, though the association of either with the SNR is still uncertain. At radio energies, the SNR has a shell-like morphology with very weak emission in the South. Using observations from the Australia Telescope Compact Array, \citet{2003MNRAS.339.1048D} concluded that the bright Western region of the SNR is most likely a PWN. They also made observations of the 21-cm H\,{\sc i} line and produced absorption spectra to obtain a lower limit on the distance of the SNR of 6 kpc. More recently, \cite{Ranasinghe_2022} re-interpreted these absorption spectra and suggested that the SNR is instead located at a distance of $3.5 \pm 0.5$ kpc. This is in better agreement with the distance of $4.41 \pm 0.50$~kpc derived by \cite{2020A&A...639A..72W} according to the additional extinction with respect to the surrounding areas traced by the near-IR photometric data from the surveys of the VISTA Variables in The Via Lactea. This SNR is also located close to a maser line emission detected at 1720 MHz with the Parkes 64-m Telescope, which is evidence of interaction with a molecular cloud \citep{1996AJ....111.1651F}, making G312.4$-$0.4 an ideal candidate for gamma-ray emission produced by proton-proton interactions.\\
This crowded region was intensively explored in the past due to the detection of the bright and unidentified EGRET source 3EG J1410$-$6147 \citep{1999ApJ...521..246C, Torres_2001, 2003MNRAS.342.1299K}. The speculation by \cite{2003MNRAS.339.1048D} that a hidden energetic pulsar might be powering the gamma-ray source was confirmed by \cite{2008MNRAS.388L...1O} who reported the discovery of the young, highly energetic pulsar PSR J1410$-$6132 located in the error box of the $\gamma$-ray source 3EG J1410$-$6147. This pulsar was detected by \emph{Fermi} just a few years after launch \citep{hou2011tracking}. It should be noted that the $\gamma$-ray pulsar discovered in blind frequency searches, PSR J1413$-$6205 \citep{2010ApJ...725..571S}, might also be responsible for part of the EGRET emission though it falls slightly outside the EGRET 95\% statistical error circle. Despite intensive effort, the energetic pulsars within the SNR, PSR~J1412$-$6145 and PSR~J1413$-$6141, have not been detected by the \emph{Fermi}-LAT \citep{Smith_2019}.\\
In this paper, we carried out a complete analysis of this extended and confused region with \emph{Fermi}-LAT data accumulated over 14 years using the most recent instrument response functions (IRFs) and diffuse background models. In Section 2, we describe the gamma-ray observations used, while Section 3 and 4 present the results obtained from a detailed morphological and spectral analysis of the LAT data. Finally, in Section 5, we discuss the main implications of these results concerning the origin of the detected gamma-ray signal.\\

\begin{figure*}[ht!]
\includegraphics[width=0.33\textwidth]{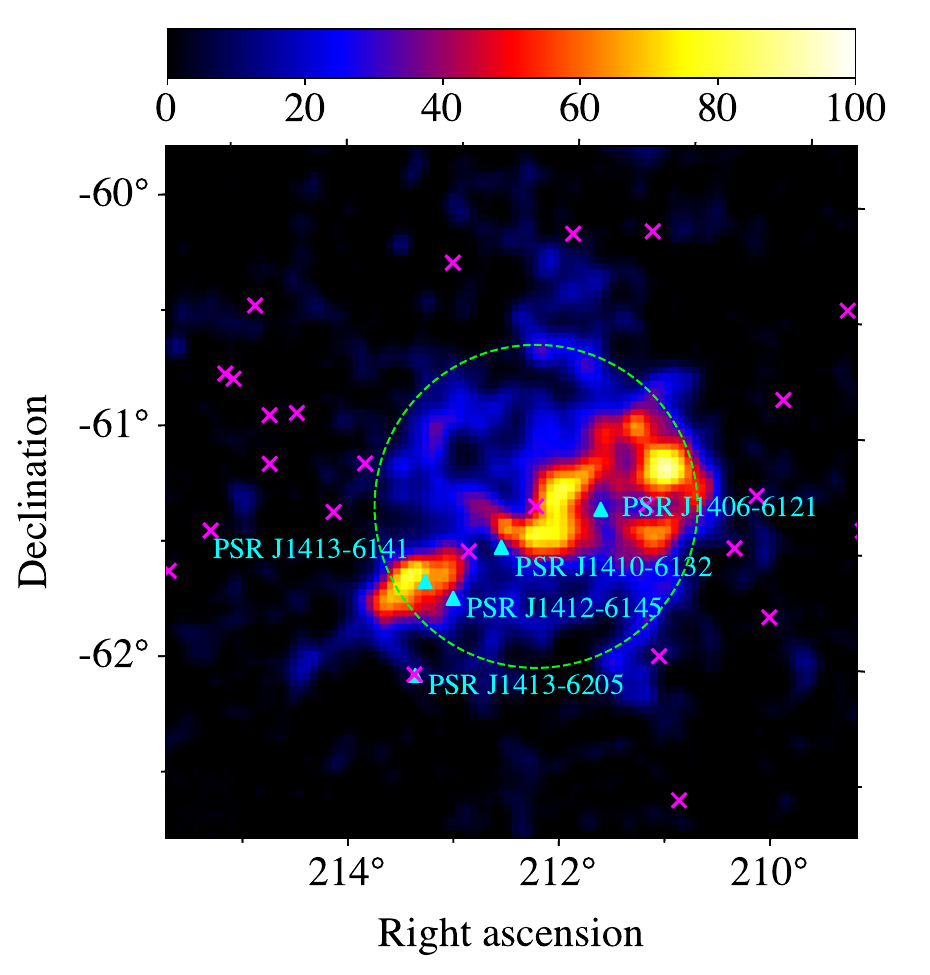}
\includegraphics[width=0.33\textwidth]{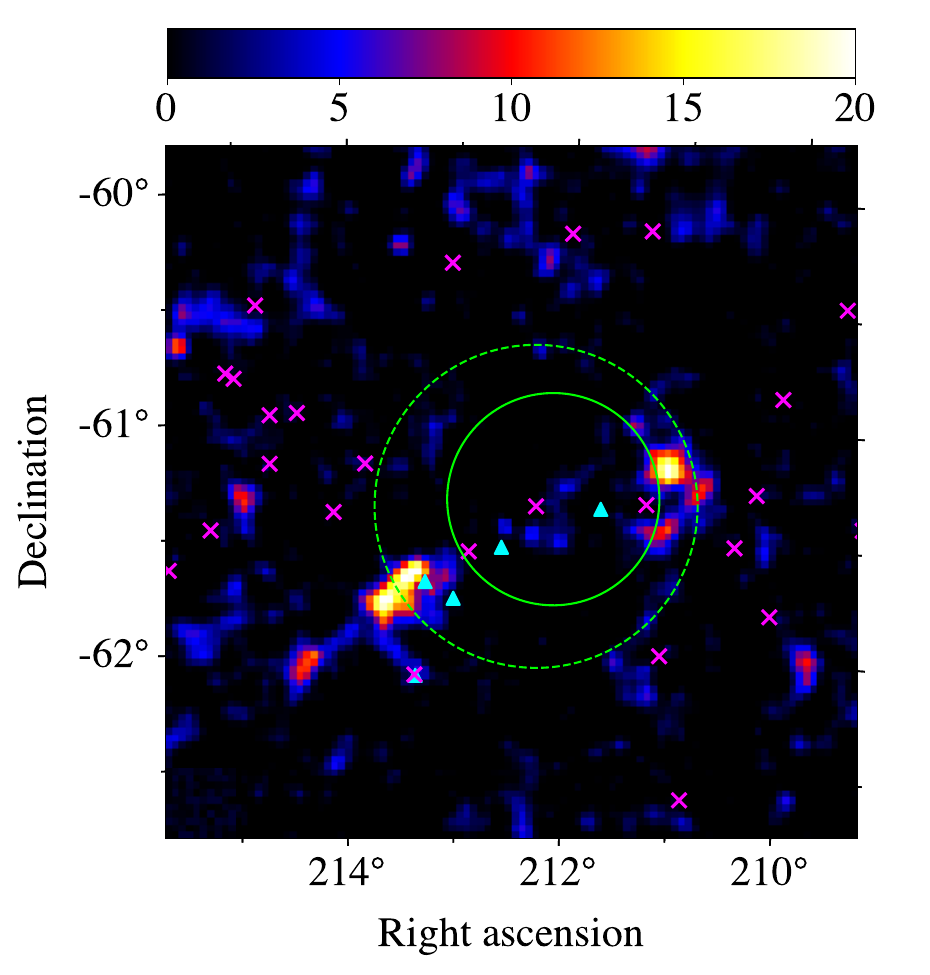}
\includegraphics[width=0.33\textwidth]{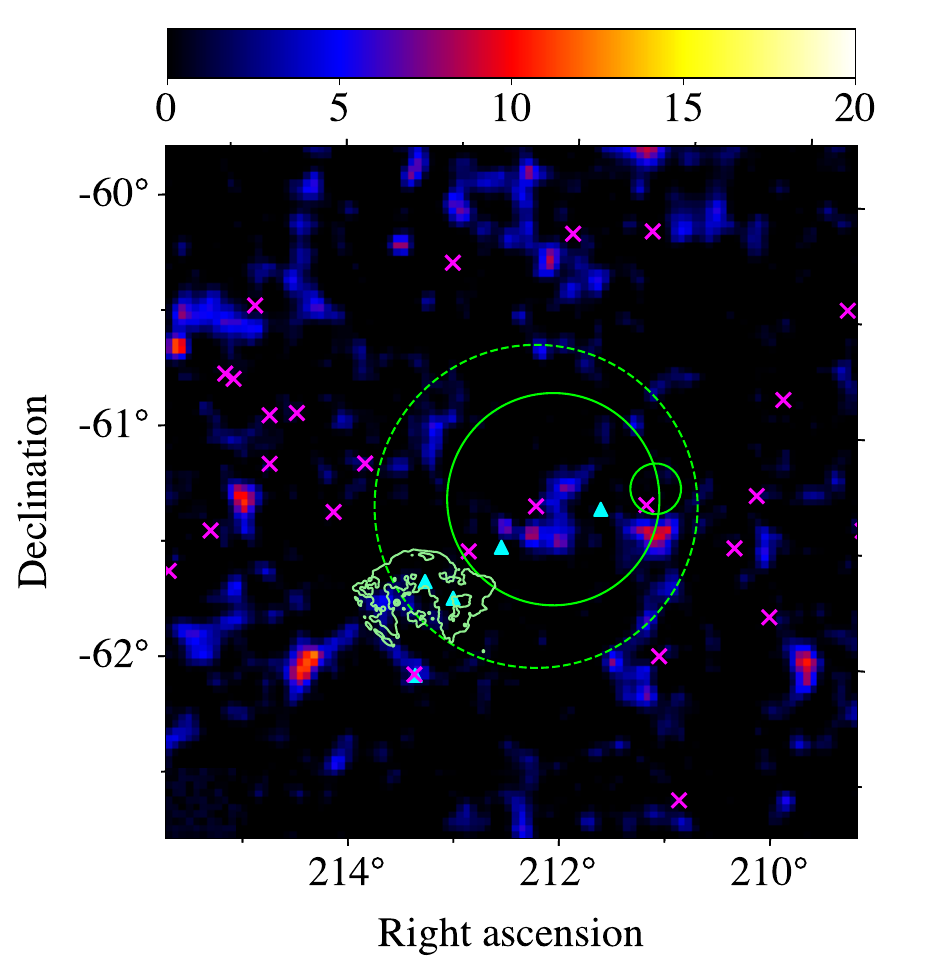}
\caption{Left panel: \emph{Fermi}-LAT TS map above 10 GeV after removing 4FGL J1409.1$-$6121e from the model. Middle panel: \emph{Fermi}-LAT TS map including 4FGL J1409.1$-$6121e as a Gaussian spatial model. Right panel: \emph{Fermi}-LAT TS map including in the model 4FGL J1409.1$-$6121e as a Gaussian, SNR G312.4$-$0.4 with the MOST template and an additional Gaussian in the North of the region. For all TS maps, we present the $3.0 \degr \times 3.0 \degr$ region of interest around 4FGL J1409.1$-$6121e above 10 GeV. MOST radio contours are in light green. The $\sigma$ extension of the best fit Gaussians are indicated with green solid lines, while the radius of the disk representing 4FGL J1409.1$-$6121e in the FGES Catalog is indicated with a green dashed circle. The magenta cross symbols represent the sources from the 4FGL-DR3 catalog and the cyan diamonds the known ATNF pulsars with $\dot{E} > 10^{35}$ erg s$^{-1}$ in the region.\\}
\label{fig:tsmaps}
\end{figure*}

\section{LAT observations} \label{sec:lat}
The \emph{Fermi}-LAT is a $\gamma$-ray telescope which detects photons by converting them into electron-positron pairs in the range from 20 MeV to higher than 500 GeV \citep{2009ApJ...697.1071A}. The following analysis is performed between 300 MeV and 3 TeV using 14 years of \emph{Fermi}-LAT data (2008 August 04 -- 2022 August 03) centered on the extended source 4FGL J1409.1$-$6121e. The current version of the LAT data is P8R3 \citep{2013arXiv1303.3514A, 2018arXiv181011394B}. We use the SOURCE class event selection, with the IRFs P8R3\_SOURCE\_V3. A maximum zenith angle of 100$^{\circ}$ between 300 MeV and 3 GeV, and 105$^{\circ}$ above 3 GeV is applied to reduce the contamination of the Earth limb. The time intervals during which the satellite passed through the South Atlantic Anomaly are excluded. Our data are also filtered removing time intervals around solar flares and bright GRBs. The data reduction and exposure calculations are performed using the LAT $fermitools$ version 2.2.0 and $fermipy$ \citep{2017ICRC...35..824W} version 1.2. Results are also cross-checked with $gammapy$ (version 0.20), an open source python library for $\gamma$-ray astronomy \citep{deil2017gammapy}. \\
We perform a binned likelihood analysis and account for the effect of energy dispersion (when the reconstructed energy differs from the true energy) by setting the parameter ${\rm edisp\_bins}=-2$. As such, the energy dispersion correction operates on the spectra with two extra bins below and above the threshold of the analysis\footnote{The energy dispersion correction is applied to all sources in the model, except for the isotropic diffuse emission model. More details can be found in the FSSC: \url{https://fermi.gsfc.nasa.gov/ssc/data/analysis/documentation/Pass8_edisp_usage.html}}. Our binned analysis is performed with 10 energy bins per decade. The Galactic diffuse emission was modeled by the standard file gll\_iem\_v07.fits and the residual background and extragalactic radiation were described by an isotropic component with the spectral shape in the tabulated model iso\_P8R3\_SOURCE\_V3\_PSF(3/2/1)\_v1.txt\footnote{\url{https://fermi.gsfc.nasa.gov/ssc/data/access/lat/BackgroundModels.html}}. \\
Since the point spread function (PSF) of the \emph{Fermi}-LAT is energy dependent and broad at low energy, we started the morphological analysis at 10 GeV while the spectral analysis was performed from 300 MeV up to 3 TeV. The minimum energy was also chosen to avoid contamination from the two bright Fermi pulsars, PSR J1410$-$6132 and PSR J1413$-$6205, located at $0.2^{\circ}$ and $0.9^{\circ}$ respectively from the centre of the analysis region. 

\begin{table}
\caption{Results of the fit of the LAT data between 10 GeV and 3 TeV using different spatial models. The second column reports the likelihood values obtained for each spatial model, while column 3 indicates the number of degrees of freedom adjusted in the model. The delta Akaike criterion, defined as $\Delta$AIC = AIC$_{disk}$ - AIC$_i$ = $2 \times (\Delta$ d.o.f - $\Delta \rm{\ln \mathcal{L}})$, is reported in the fourth column. See Sect.~\ref{sec:morpho} for more details.}
\label{tab:aic} 
\centering    
\begin{tabular}{lccc}
\hline \hline
Spatial model (number) & Likelihood & d.o.f & $\Delta$AIC \\
\hline
Disk        (1) &  $-$83832.1  & 5 & 0 \\
Gaussian (2)      &  $-$83819.2   & 5 & 25.8 \\
Gaussian + MOST template  (3) &  $-$83804.1 & 7 &  52.0 \\
Gaussian + MOST template + Point source (4)& $-$83792.6   & 11 & 67.0 \\
Gaussian + MOST template + Gaussian (5) & $-$83784.6   & 12 &  81.0 \\
Gaussian + MOST template with wing + Gaussian (6) & $-$83787.0   & 12 & 76.2 \\
Gaussian + SNR disk + Gaussian (7) & $-$83785.2  & 15 & 73.8 \\
\end{tabular}
\end{table}

\section{Morphological analysis} \label{sec:morpho}
The morphological analysis was performed through a binned analysis using all event types with spatial bins of $0.03^{\circ}$ over a region of $9^{\circ} \times 9^{\circ}$. We included all sources from the LAT 12-year Source Catalog (4FGL-DR3)\footnote{\url{https://fermi.gsfc.nasa.gov/ssc/data/access/lat/12yr_catalog}} in a region of $15^{\circ} \times 15^{\circ}$. As a first step, the spectral parameters of the sources located up to $3^{\circ}$ from the center of the region of interest (ROI) were fit simultaneously with the Galactic and isotropic diffuse emissions. During this procedure, the 4FGL-DR3 log-parabola spectral model of 4FGL J1409.1$-$6121e was used to reproduce the $\gamma$-ray emission of the extended source fixing its spatial model to the disk of radius $0.73^{\circ}$ published in~\cite{2017ApJ...843..139A}. To search for additional sources in the ROI, we computed a test statistic (TS) map that tests at each pixel the significance of a source with a generic E$^{-2}$ spectrum against the null hypothesis: ${\rm TS}=2(\ln \mathcal{L}_1 - \ln \mathcal{L}_0)$, where $\mathcal{L}_0$ and $\mathcal{L}_1$ are the likelihoods of the background (null hypothesis) and the hypothesis being tested (source plus background). We iteratively added two point sources in the model where the TS exceeded 25. We localized the two additional sources (RA$_{\rm J2000}$, Dec$_{\rm J2000}$ = $205.36^{\circ}, -63.63^{\circ}$; $215.14^{\circ}, -64.39^{\circ}$) and we fitted their power-law spectral parameters.\\ 
We then performed the morphological analysis of 4FGL J1409.1$-$6121e whose results are reported in Table~\ref{tab:morpho}. In each step, we fit both the morphological and spectral parameters of the new components, the source associated with 4FGL J1409.1$-$6121e being fit with a logarithmic parabola spectral model\footnote{The logarithmic parabola spectral model is defined as: 
$\frac{dN}{dE} = N_0 \left( \frac{E}{E_0} \right)^{-\left(\alpha + \beta \log\left(E/E_0\right) \right)}$} while all other new components are fit with a simple power-law model. When using geometrical templates, both the position and the angular size of the sources of interest are being fit. The different steps of the procedure are the following: fitting a disk for the source 4FGL J1409.1$-$6121e, replacing the disk by a Gaussian source (middle panel of Figure~\ref{fig:tsmaps}), generating and adding a custom MOST spatial template of the SNR G312.4$-$0.4 in addition to the Gaussian of $0.46^{\circ}$ now modeling 4FGL J1409.1$-$6121e and finally adding a second Gaussian to the model hereafter referred to as Gauss2 of size $0.11^{\circ}$. This last step provides the best fit to the data, as can be seen in Figure~\ref{fig:tsmaps} (right). From Table~\ref{tab:aic}, it can be concluded that extended emission spatially associated with G312.4$-$0.4 is significantly detected. This is assessed quantitatively using the likelihood ratio test, i.e., through the values of TS = 2 $\times \Delta \rm{\ln \mathcal{L}} = 30.2$ ($5.1\sigma$ for two additional degrees of freedom) for the model using the radio template compared to the hypothesis of no emission associated with G312.4$-$0.4. This is also clearly observed in the excess TS map (Figure~\ref{fig:g312} (left)), highlighting the excellent correlation between the gamma-ray emission and the radio contours. Since we cannot use the likelihood ratio test to compare models that are not nested, we used the Akaike Information Criterion \cite[AIC]{Akaike1998}. We calculated $\Delta$AIC = AIC$_{disk}$ - AIC$_i$ = $2 \times (\Delta$ d.o.f - $\Delta \rm{\ln \mathcal{L}})$ to compare the AIC of the simple disk hypothesis and the AIC of the more complex models. The best $\Delta$AIC value, reported in this Table, is obtained for the radio template from MOST, excluding the Western wing which is most likely associated with a PWN~\citep{2003MNRAS.339.1048D}. This is supported by the fact that the more complex spatial models 6 and 7 (where we add the Western wing to the radio template or replace the radio template by a disk hypothesis) do not improve significantly the fit. Table~\ref{tab:morpho} summarizes the morphological parameters of our best fit model; the extended source representing the emission of 4FGL J1409.1$-$6121e is kept at its original name, though the best-fit centroid is slightly shifted in comparison to the initial value (RA$_{\rm J2000}$, Dec$_{\rm J2000}$ = $212.29^{\circ}, -61.353^{\circ}$).\\

\section{Spectral analysis} \label{sec:spec}
\begin{figure*}[ht!]
\includegraphics[width=0.45\textwidth]{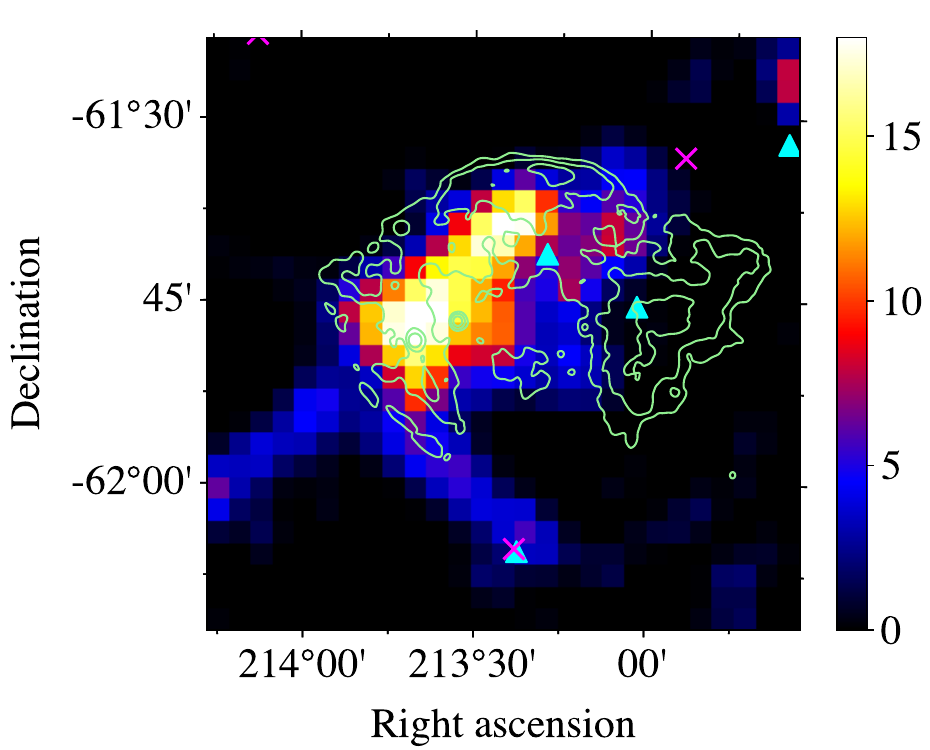}
\hspace{0.5cm}
\includegraphics[width=0.52\textwidth]{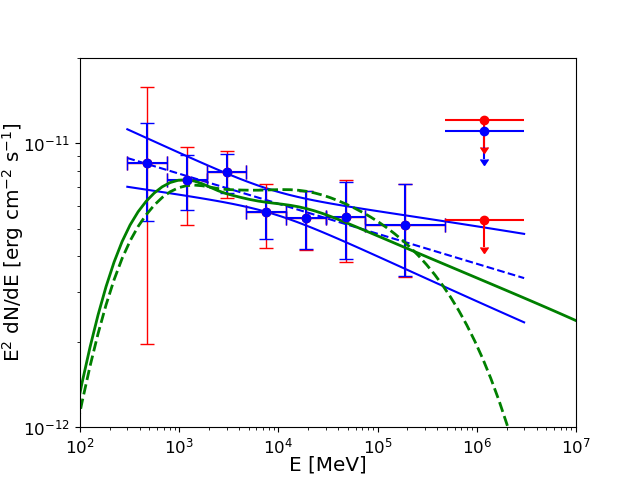}
\caption{Left panel: Zoom on the \emph{Fermi}-LAT TS map at the position of the SNR G312.4$-$0.4 above 10 GeV in equatorial coordinates (the SNR is not included in the model). MOST radio contours are indicated in light green. Magenta cross symbols represent the sources from the 4FGL-DR3 catalog and the cyan diamonds the known ATNF pulsars with $\dot{E} > 10^{35}$ erg s$^{-1}$ in the region.
Right panel: Spectral energy distribution of the SNR G312.4$-$0.4 obtained using the MOST spatial template. Blue error bars represent the statistical uncertainties, while the red ones correspond to the statistical and systematic uncertainties added in quadrature. In the last energy bin, the two red arrows indicate the extrema of upper limits obtained with the different systematics. Hadronic models considering a proton distribution following a pure power law of index 2.2 (green line) or power-law of index 2.1 with an exponential cutoff at 10 TeV (green dotted line) are superimposed.}
\label{fig:g312}
\end{figure*}

Using the best-fit spatial template (Model 5 in Table~\ref{tab:aic}), we performed the spectral analysis from 300 MeV to 3 TeV over a region of $15^{\circ} \times 15^{\circ}$, including all sources from the 4FGL-DR3 Catalog over a region of $25^{\circ} \times 25^{\circ}$. The summed likelihood method was used to simultaneously fit events with different angular reconstruction quality (PSF0 to PSF3 event types\footnote{\url{https://fermi.gsfc.nasa.gov/ssc/data/analysis/documentation/Cicerone/Cicerone_Data/LAT_DP.html}}). Here, we used PSF1, PSF2 and PSF3 events below 3 GeV with spatial bins of $0.15^{\circ}$, and all event types above 3 GeV with spatial bins of $0.05^{\circ}$. Since we used two additional years of data with respect to the 4FGL-DR3 Catalog, we first checked whether additional sources were needed in the model by examining the TS maps above 300 MeV. Six additional sources were detected at RA$_{\rm J2000}$, Dec$_{\rm J2000}$ = $197.96^{\circ}, -62.72^{\circ}$; $203.80^{\circ}, -62.61^{\circ}$; $205.23^{\circ}, -61.84^{\circ}$; $205.50^{\circ}, -62.42^{\circ}$; $212.57^{\circ}, -59.31^{\circ}$; $225.67^{\circ}, 59.07^{\circ}$ (not detected in the 10 GeV -- 3 TeV range used in Sect.~\ref{sec:morpho} but all located more than $2^{\circ}$ from the center of our region of interest). We then tested a simple power-law model and a logarithmic parabola for the three extended sources detected in our Model 5. During this procedure, the spectral parameters of sources located up to $3^{\circ}$ from the ROI center were again left free during the fit, like those of the Galactic and isotropic diffuse emissions. The improvement between the power-law model and the logarithmic parabola was tested using the likelihood ratio test. It is significant only in the case of the large Gaussian source coincident with 4FGL J1409.1$-$6121e.\\ 
The systematic errors on the morphological and spectral analysis depend on the uncertainties on the Galactic diffuse emission model, on the effective area, and on the spatial shape of the source. The first is calculated using eight alternative diffuse emission models following the same procedure as in the first \emph{Fermi}-LAT supernova remnant catalog  \citep{2016ApJS..224....8A}  and the second is obtained by applying two scaling functions on the effective area following the standard method defined in \cite{2012ApJS..203....4A}. We consider the impact on the spectral parameters when changing the spatial model from the MOST template to the best disk hypothesis for the SNR. These three sources of systematic uncertainties were added in quadrature to represent the total systematic uncertainty. The spectral parameters of the three sources are presented in Table~\ref{tab:spec}, together with their estimated systematic errors. As one can see in this Table, the low TS value of the gamma-ray source associated with the SNR and with Gauss2 prevents us from doing any search of variability on timescales smaller than one year. This search was performed for the three sources and no variability on a 1-year timescale was detected. \\
We finally derived the \emph{Fermi}-LAT spectral points for the SNR G312.4$-$0.4, shown in Fig.~\ref{fig:g312} (right panel), by dividing the 300 MeV -- 3 TeV energy range into 10 logarithmically-spaced energy bins and  performing a maximum likelihood spectral analysis to estimate the photon flux in each interval, assuming a power-law shape with fixed photon index $\Gamma$=2  for the source. The normalizations of the diffuse Galactic and isotropic emission were left free in each energy bin as well as those of the sources within $3^{\circ}$. A 95\% C.L. upper limit is computed when the TS value is lower than $1$. The four energy bins above 100 GeV were combined in two logarithmically-spaced bins to reduce the statistical uncertainties.

\begin{table}
\caption{\emph{Fermi}-LAT morphological parameters of the two Gaussian sources in Model 5 derived between 10 GeV and 3 TeV.}
\label{tab:morpho}
\centering    
\begin{tabular}{lcccc}
\hline \hline
Source name & RA & Dec & Gaussian $\sigma$ & TS \\
 & ($^{\circ}$) & ($^{\circ}$) &  \\ 
\hline
4FGL J1409.1$-$6121e      &   $212.14$ & $-61.32$  & $0.46$ & 5351 \\
Gauss2     & $211.22$ & $-61.26$  & $0.11$ & 64 \\
\end{tabular}
\end{table}

\begin{table}
\caption{\emph{Fermi}-LAT spectral parameters of the extended sources in Model 5 between 300 MeV and 3 TeV. The third column provides the curvature parameter $\beta$ used for the log-parabola model. The first (second) error represent statistical (systematic) error respectively. Columns 5 and 6 provide the TS value and the improvement of the log-normal representation with respect to the PL model TS$_{LP}$.}
\label{tab:spec}
\centering    
\begin{tabular}{lccccc}
\hline \hline
Source name & Spectral index & $\beta$ & Energy flux (300 MeV - 3 TeV) & TS & TS$_{LP}$ \\
 & & &  ($10^{-11}$ erg cm$^{-2}$ s$^{-1}$) & \\ 
\hline
G312.4$-$0.4        & $2.10 \pm 0.05 \pm 0.04$ &  & $2.8 \pm 0.3 \pm 0.2$ &  177 & 0.1 \\
4FGL J1409.1$-$6121e       & $1.96 \pm 0.03 \pm 0.03$ & 0.60 $\pm 0.07 \pm 0.02$ & $25.4 \pm 0.5 \pm 0.9$ &  5370 & 65.2 \\
Gauss2      & $1.86 \pm 0.06 \pm 0.05$ & & $2.5 \pm 0.4 \pm 0.7$ & 64  & 2.9 \\
\end{tabular}
\end{table}

\section{Discussion} \label{sec:discus}
Using 14 years of LAT data and a more precise morphological and spectral analysis, we have been able to significantly detect three sources in the immediate region of the extended source 4FGL J1409.1$-$6121e. The brightest one, coincident with 4FGL J1409.1$-$6121e, was only slightly affected by our increased statistics and re-analysis since its spectral parameters are consistent with those of its initial detection but its centroid is slightly shifted, in addition to using a Gaussian of smaller size of $0.46^{\circ}$ (with respect to $0.51^{\circ} \pm 0.02^{\circ}$, see Table~3 from \citet{2017ApJ...843..139A}). The other main differences are the detection of a new source whose gamma-ray emission correlates well with the radio shell of the SNR G312.4$-$0.4, and the detection of a second, smaller Gaussian source which lies in the vicinity of the binary system 4FGL J1405.1$-$6119.\\
The gamma-ray luminosity of the large Gaussian source coincident with 4FGL J1409.1$-$6121e corresponds to $2.36 \times 10^{34} (\rm{D/1 kpc})^{2}$ erg s$^{-1}$. This can be compared to the rotational energy of the two closest pulsars: $1.0 \times 10^{37}$ erg s$^{-1}$ for PSR J1410$-$6132 located at 13.51 kpc and $2.2 \times 10^{35}$ erg s$^{-1}$ for PSR J1406$-$6121 located at 7.3 kpc~\citep{2005AJ....129.1993M}. The very high rotational energy of PSR J1410$-$6132 would be sufficient to power the gamma-ray source. At the assumed distance of the pulsar of 13.51 kpc, the gamma-ray extension of the source would imply a physical extension of 108 pc. This demonstrates that 4FGL J1409.1$-$6121e is certainly confused as was already suggested by \citet{2017ApJ...843..139A}. 
In the same way, the gamma-ray luminosity of the small Gaussian source, located at $0.06^{\circ}$ from the binary system 4FGL J1405.1$-$6119, corresponds to $2.82 \times 10^{33} (\rm{D/1 kpc})^{2}$ erg s$^{-1}$. This high luminosity is unlikely to be powered by the close-by pulsar PSR J1406$-$6121 alone since it would require an efficiency of conversion of spin-down power into gamma rays of $\sim 50$\%. The gamma-ray flux could also be overestimated due to contamination from a probable point-like inverse Compton emission produced by the binary system 4FGL J1405.1$-$6119, as it is the case in other gamma-ray binaries (see \cite{2013A&ARv..21...64D} for a review). However, the extension of the source demonstrates that a different component dominates at gamma-ray energies since extension is not expected for gamma-ray binaries. Interestingly, the 2MASS catalogue \citep{2006AJ....131.1163S} lists the long-period variable source 2MASS J14045171$-$6115578 which lies only 23 arcsec away from the position of the $\gamma$-ray source, providing another possible origin of the signal. As for the case of the binary system 4FGL J1405.1$-$6119, such scenario would not explain the extended signal detected by \emph{Fermi}.\\
The two Gaussian sources, 4FGL J1409.1$-$6121e and Gauss2, thus remain unidentified and, among the three extended sources in this region, only the one coincident with the SNR can be safely associated thanks to the connection found with the MOST radio contours.\\ 
The gamma-ray luminosity of the source coincident with the SNR G312.4$-$0.4, $1.80 \times 10^{33} (\rm{D/1 kpc})^{2}$ erg s$^{-1}$ between 1 GeV and 100 GeV, is among the faintest reported SNRs detected by Fermi-LAT as can be seen in \cite{2016ApJS..224....8A}. This value is close to the luminosity of the 4 kyr SNR Puppis A in which a shock-cloud interaction was reported in the eastern side by \cite{hfp05}: $5.58 \times 10^{33} (\rm{D/1 kpc})^{2}$ erg s$^{-1}$ \citep{2012ApJ...759...89H}. Interestingly, the spectral shape reported in this {\emph Fermi}-LAT detection paper\footnote{A re-analysis of the SNR with more data was able to detect the presence of a break at 8 GeV \citep{2017ApJ...843...90X}.} is similar to the one reported here with a pure power-law of index $\sim 2.1$. \cite{2012ApJ...759...89H} and \cite{2017ApJ...843...90X} have shown that, although the inverse Compton-dominated and bremsstrahlung-dominated models can marginally explain the $\gamma$-ray emission, the hadron-dominated model seems to be more plausible since the first two scenarios require electron-to-proton ratios higher than 0.1 (to reduce emission from proton-proton interaction), in excess of cosmic-ray abundances. The same conclusion would apply here due to the similar energetics at $\gamma$-ray energies.\\
We therefore focused on the hadronic scenario to derive constraints from the gamma-ray data. The leptonic scenario cannot be excluded but due to the lack of multi-wavelength data for the source, in particular the X-ray, we cannot draw firm conclusions. The eROSITA SNR catalog may be able to provide additional information in the future. We employed the Naima package~\citep{naima} to model the gamma-ray spectrum, with a single population for simplicity. In reality, emission may arise from a number of regions, such as the main shell of the shocked ambient medium, or from re-acceleration of pre-existing cosmic rays by radiative shocks in the adjacent clouds.\\ 
The best-fit obtained assuming a proton distribution following a pure power-law of index 2.2 or a power law of index 2.1 with an exponential cutoff at 10 TeV are superimposed on the gamma-ray spectral points in Figure~\ref{fig:g312} (right). Both models require a total CR proton energy W ($>1$ GeV) $\sim 5 \times 10^{50} (\rm{n/1 cm^{-3}})$ erg. This implies that distances larger than 4~kpc are disfavored, in the hadronic scenario, unless the medium is very dense. Interestingly, the catalog of molecular clouds in the Milky Way by \cite{2016ApJ...822...52R} reports the presence of a dense cloud at a distance of 3.68~kpc, with a total mass of $4.5 \times 10^4 \, \rm{M_{\odot}}$ and a radius of $14.42$~pc, implying an average density of 103 cm$^{-3}$ and an energy injected into protons of only 0.5\% of the kinetic energy of a supernova. This energy estimate should be treated as a lower limit since only part of the shock may be interacting with the molecular cloud. Combining this with our new gamma-ray data, one would favor a distance of 3.7~kpc for this SNR. At this distance, the SNR would have a diameter of 25.6~pc (excluding the Western wing probably associated with an unrelated PWN). Assuming a kinetic energy of $10^{51}$ erg for the SN and an average density of the medium of 1 cm$^{-3}$, this physical extension would lead to an age of 11~kyrs using the expression for the dynamic age of an SNR in the Sedov expansion phase from \cite{1999ApJ...521..246C}. Despite the lack of spectral cut-off detected at gamma-ray energies, SNR G312.4$-$0.4 would be a medium-age SNR, much older than X-ray bright SNRs such as Cas A, Kepler, or Tycho, and more similar to the transition SNR Puppis A.\\
Deeper multi-wavelength observations are required to better constrain the distance and the environment of these three sources. In particular, SNR G312.4$-$0.4 is a good candidate to be observed with the next generation of Cherenkov telescopes CTA \citep[Cherenkov Telescope Array,][]{2019scta.book.....C} that will give clear insights into the maximum particle energy. In turn, these TeV observations with increased sensitivity and spatial resolution will help to constrain the nature of the two unidentified Gaussian sources detected with our improved morphological and spectral analysis.

\begin{acknowledgments}
The \textit{Fermi} LAT Collaboration acknowledges generous ongoing support
from a number of agencies and institutes that have supported both the
development and the operation of the LAT as well as scientific data analysis.
These include the National Aeronautics and Space Administration and the
Department of Energy in the United States, the Commissariat \`a l'Energie Atomique
and the Centre National de la Recherche Scientifique / Institut National de Physique
Nucl\'eaire et de Physique des Particules in France, the Agenzia Spaziale Italiana
and the Istituto Nazionale di Fisica Nucleare in Italy, the Ministry of Education,
Culture, Sports, Science and Technology (MEXT), High Energy Accelerator Research
Organization (KEK) and Japan Aerospace Exploration Agency (JAXA) in Japan, and
the K.~A.~Wallenberg Foundation, the Swedish Research Council and the
Swedish National Space Board in Sweden.
 Additional support for science analysis during the operations phase is gratefully
acknowledged from the Istituto Nazionale di Astrofisica in Italy and the Centre
National d'\'Etudes Spatiales in France. This work performed in part under DOE
Contract DE-AC02-76SF00515.\\
PC and MLG acknowledge ANR for support to the GAMALO project under reference ANR-19-CE31-0014.\\
AS acknowledges support from the Spanish Ministry of Universities through the Maria Zambrano Talent Attraction Programme, 2021--2023.
\end{acknowledgments}

%


\bibliography{bibliography_g312}{}
\bibliographystyle{aasjournal}



\end{document}